\documentclass[11pt]{article}
\usepackage[dvips]{graphics}
\title{Covariant many-fingered time Bohmian interpretation 
of quantum field theory}
\author{Hrvoje Nikoli\'c \\
Theoretical Physics Division, Rudjer Bo\v{s}kovi\'{c} Institute, \\
P.O.B. 180, HR-10002 Zagreb, Croatia \\
{\normalsize e-mail: hrvoje@thphys.irb.hr} \\
\makebox[1in]{} \\
}
\date{\today}
\begin{document}
\maketitle
\begin{abstract}
The Bohmian interpretation of the many-fingered time (MFT)
Tomonaga-Schwinger formulation of quantum field theory (QFT)
describes MFT fields, which provides a
covariant Bohmian interpretation of QFT without introducing
a preferred foliation of spacetime.
\end{abstract}

\vspace*{0.5cm}


\maketitle

\section{Introduction}

The Bohmian deterministic interpretation of quantum field theory 
(QFT) \cite{bohm2,bohmrep,holrep,holbook,nikfpl1} 
is a promising approach to the general solution of the problem
of measurement in QFT.
A particular advantage of this interpretation is the fact that it
automatically solves the problem of time in canonical quantum gravity
\cite{holrep,gol,pin1,pin2,shoj}.
In addition, this interpretation
might play an important role in some areas of high-energy physics,
such as quantum cosmology \cite{bar,col,pn,mar,pn2}
and noncommutative theories \cite{barb}.
An important problem of this interpretation
is the fact it requires a preferred foliation of spacetime,
which makes it noncovariant \cite{holrep,shoj}. The source
of this problem can be traced back to the fact that the
functional Schr\"odinger equation of QFT 
is also noncovariant because it also requires
a preferred foliation of spacetime. 

The problem of noncovariance of the Schr\"odinger equation
can be solved by replacing the usual
time-dependent Schr\"odinger equation with the
{\em many-fingered time} (MFT)
Tomonaga-Schwinger equation \cite{tomo,schw}, which does not
involve a preferred foliation of spacetime. In this formulation,
the quantum state is a functional of an arbitrary
timelike hypersurface. In a manifestly covariant formulation
introduced in \cite{dopl}, the hypersurface does not even need to be
timelike.

The covariance of the Tomonaga-Schwinger equation
suggests the possibility to formulate a covariant Bohmian
interpretation based on the Tomonaga-Schwinger equation,
rather than on the Schr\"odinger equation. Indeed, such an attempt
has recently been performed in \cite{hort}. 
(For other recent approaches to the covariant Bohmian formulation
of relativistic quantum mechanics and QFT, see also
\cite{nikrqm,nikDW}).
However, the attempt in \cite{hort} was based on
a Bohmian equation of motion with a {\em single} time, 
so some extra rule that determined the foliation was still
necessary. 
In contrast with \cite{hort}, in  
this paper we observe that a natural
Bohmian interpretation of the MFT Tomonaga-Schwinger equation
involves a MFT Bohmian equation of motion,
which avoids the need for an extra rule that determines the 
foliation. 

The paper is organized as follows. After reviewing the MFT 
formulation of QFT in Sec.~\ref{MFT}, we discuss the MFT Bohmian interpretation 
in Sec.~\ref{BOHM}.
In Sec.~\ref{COV} we demonstrate that all results of this paper 
can be written in a manifestly covariant form, which
leads to a covariant form of the Bohmian interpretation,
without involving a preferred foliation of spacetime.

For simplicity, in this paper we study only a real
scalar field. However, 
it is straightforward to apply the developed MFT formalism
to a Bohmian interpretation of any other quantum field,
including fermionic fields \cite{nikfpl2} and the gravitational field
\cite{holrep,gol,pin1,pin2,shoj}.

In the paper we use units $\hbar=c=1$, while the signature of 
spacetime metric is $(+,-,-,-)$.

\section{MFT formulation of QFT}
\label{MFT}

Let $x=\{x^{\mu}\}=(x^0,{\bf x})$ be spacetime coordinates.
A timelike Cauchy hypersurface $\Sigma$ can be defined by a function
$T({\bf x})$, through the equation
\begin{equation}\label{param1}
x^0=T({\bf x}).
\end{equation}
The coordinates ${\bf x}$ may be used as coordinates on $\Sigma$.
If $T({\bf x})$ is given for all ${\bf x}$, then
${\bf x}$ can also be viewed as a point on $\Sigma$ and allows 
us to write ${\bf x}\in\Sigma$.
Let $\phi({\bf x})$ be a dynamical field on $\Sigma$.
(For simplicity, we take it to be
a real scalar field.) 
To avoid notational confusion,
from now on $\phi({\bf x})$ and $T({\bf x})$ denote a value
at a point ${\bf x}$,
while $\phi$ and $T$ without an argument 
denote a set of values at {\em all}
points ${\bf x}$. 
Similarly, if $\sigma$ is a subset 
of $\Sigma$, then $\phi|_{\sigma}$ and $T|_{\sigma}$ denote a
set of values at all points of ${\sigma}$. (With this notation,
$\phi=\phi|_{\Sigma}$, $T=T|_{\Sigma}$.)
Let $\hat{{\cal H}}({\bf x})$ be the 
Hamiltonian-density operator.
The dynamics of the field $\phi$ is described by 
the MFT Tomonaga-Schwinger equation 
\begin{equation}\label{TS}
\hat{{\cal H}}({\bf x})\Psi[\phi,T]=
i\frac{\delta \Psi[\phi,T] }{\delta T({\bf x})}.
\end{equation}
The wave functional $\Psi[\phi,T]$ 
can also be viewed as a functional of $\phi|_{\Sigma}$, 
where $\Sigma$ is defined by $T$.
Eq.~(\ref{TS}) 
describes how $\Psi$ changes for an infinitesimal 
change $\delta T({\bf x})$ of the hypersurface $\Sigma$. 
One easily recovers the usual 
functional Schr\"odinger equation by integrating 
(\ref{TS}) over the entire hypersurface $\Sigma$, 
provided that one considers a special form of the 
variation $\delta T({\bf x})$, such that $\delta T({\bf x})$
does not depend on ${\bf x}$. Thus we see that (\ref{TS})
represents a generalization of the ordinary Schr\"odinger equation.
However, in contrast to the ordinary Schr\"odinger equation, the 
right-hand side of (\ref{TS}) does not involve any preferred 
foliation of spacetime. We also note that, in the original papers
\cite{tomo,schw}, $\hat{{\cal H}}({\bf x})$ 
was the interaction Hamiltonian density 
in the interaction picture, while here, in accordance with
\cite{dopl,hort}, $\hat{{\cal H}}({\bf x})$ is the total Hamiltonian density 
in the Schr\"odinger picture.

The quantity $\rho[\phi,T]=|\Psi[\phi,T]|^2$ represents 
the probability density for the field to have a value $\phi$ 
on $\Sigma$. Since $\Sigma$ is determined by $T$, it is also 
convenient to say that $\rho[\phi,T]$ is the probability density for the field
to have a value $\phi$ at time $T$. However, when using the 
latter terminology, it is important to remember that $T$ is not a 
single real parameter, but a collection of an infinite 
number of real parameters, with one real parameter 
for each point ${\bf x}$. 

Now consider a quantum measurement.
It is convenient to describe
a measurement in terms
of a wave-functional collapse. The collapse can be described in a general
way as follows. A normalized solution $\Psi[\phi,T]$ of (\ref{TS})
can be written as a linear combination of other orthonormal solutions
as
\begin{equation}\label{expan}
\Psi[\phi,T]=\sum_a c_a \Psi_a[\phi,T].
\end{equation}
For any given $\Psi_a$ and any given $T_0$, there
exists a hermitian operator such that $\Psi_a[\phi,T_0]$
is an eigenvector of this operator. Since any hermitian operator
corresponds to a quantity that can, at least in principle, be measured,
a measurement at $T_0$ may induce a wave-functional collapse of the form
\begin{equation}\label{collaps}
\Psi[\phi,T] \rightarrow \Psi_a[\phi,T].
\end{equation}
The probability for the
collapse (\ref{collaps}) is $|c_a|^2$.
In particular, $\Psi_a[\phi,T]$ may be a product
of the form 
\begin{equation}\label{loc2}
\Psi_a[\phi,T]=A[\phi|_{\bar{\sigma}},T|_{\bar{\sigma}}] 
\prod_{{\bf x}\in\sigma}
\delta(\varphi({\bf x})-\phi({\bf x})) ,
\end{equation}
where $\sigma$ is a subset of $\Sigma$ and 
$\bar{\sigma}\equiv\Sigma-\sigma$ denotes 
the set of all points        
of $\Sigma$ that are not contained in $\sigma$.
The state (\ref{loc2}) has a well defined value of $\phi|_{\sigma}$.
Therefore, if
(\ref{loc2}) is satisfied, then the collapse
(\ref{collaps}) may occur by measuring $\phi|_{\sigma}$ at
$T_0|_{\sigma}$, without any measurement on
$\bar{\sigma}$.
(This is similar to the fact that the total wave function of an
entangled Einstein-Podolsky-Rosen (EPR)
pair collapses by measuring the spin of one member
of the pair, without measuring the spin of the other member.
For a MFT description of the EPR effect, see \cite{ghose}.)
In this sense, a measurement can assign a definite value of
$\phi|_{\sigma}$ on $\sigma$, without saying anything about
$T|_{\bar{\sigma}}$ and $\phi|_{\bar{\sigma}}$. In particular, 
this implies that, in the MFT formulation, the wave-functional 
collapse does not need to be ``instantaneous",
because the time of measurement $T_0|_{\sigma}$
does not determine the ``collapse time" $T_0|_{\bar{\sigma}}$ 
of the unmeasured variables.

A measurement can also be described more accurately by introducing
the degrees of freedom $\chi$ of the measuring apparatus.
If the apparatus measures the eigenvalues associated with the states
$\Psi_a$, then the total wave functional describing the entanglement
between the measured system and the measuring apparatus takes the
form
\begin{equation}\label{expanmeas}
\Psi[\phi,\chi,T]=\sum_a c_a \Psi_a[\phi,T] \Theta_a[\chi,T].
\end{equation}
Here $\Theta_a[\chi,T]$ are orthonormal functionals that do not overlap,
in the sense that
\begin{equation}\label{theta}
\Theta_a[\chi,T] \Theta_{a'}[\chi,T]=0 \;\;\; {\rm for} \;\;\;
a\neq a' .
\end{equation}
Therefore, if $\chi$ at $T_0$ is found to have
a value that belongs to the support
of $\Theta_a$, then the value of $\chi$ at $T_0$ does not belong to the
support of any other $\Theta_{a'}$. In this case, the total
effective wave functional is given by
$\Psi_a[\phi,T] \Theta_a[\chi,T]$.
Consequently, the effective wave functional describing only the
$\phi$-degree od freedom is $\Psi_a[\phi,T]$,
which corresponds to the collapse (\ref{collaps}).
The probability for this effective collapse is equal to $|c_a|^2$.
Of course, the conventional interpretation of quantum mechanics
does not say what, if anything, causes $\chi$ to take a definite value.
However, such a cause is provided by the Bohmian interpretation
discussed in the next section.

\section{Bohmian interpretation}
\label{BOHM}

For simplicity, consider a 
free scalar field with the Hamiltonian density 
\begin{equation}
\hat{{\cal H}}({\bf x}) =
-\frac{1}{2} \frac{\delta^2}{\delta\phi^2({\bf x})}
+ \frac{1}{2} [(\nabla\phi({\bf x}))^2 + m^2\phi^2({\bf x})].
\end{equation}
By writing $\Psi=Re^{iS}$, where
$R$ and $S$ are real functionals, the complex 
equation (\ref{TS}) is equivalent to a set of two real equations
\begin{equation}\label{HJ}
\frac{1}{2} \left( \displaystyle\frac{\delta S}{\delta\phi({\bf x})} 
\right)^2
+\frac{1}{2} [(\nabla\phi({\bf x}))^2 +m^2\phi^2({\bf x})] 
+{\cal Q}({\bf x};\phi,T]   
+\displaystyle\frac{\delta S}{\delta T({\bf x})} =0 ,  
\end{equation}
\begin{equation}\label{eqvar}
\frac{\delta\rho}{\delta T({\bf x})} +
\frac{\delta}{\delta\phi({\bf x})} \left(\rho
\frac{\delta S}{\delta\phi({\bf x})} \right) =0,
\end{equation}
where $\rho=R^2$ and
\begin{equation}
{\cal Q}({\bf x};\phi,T]=-\frac{1}{2R}
\frac{\delta^2R}{\delta\phi^2({\bf x})}.
\end{equation}
The conservation equation (\ref{eqvar}) shows that
it is consistent to interpret $\rho[\phi,T]$
as the probability density for the field to have the value
$\phi$ at the hypersurface determined by $T$.

The Bohmian interpretation consists in introducing a deterministic
time-dependent hidden variable, such that the time evolution of
this variable is consistent with the probabilistic interpretation
of $\rho$. From (\ref{eqvar}), we see that this is naturally
achieved by introducing a MFT field $\Phi({\bf x}; T]$ 
that satisfies the MFT Bohmian equation of motion
\begin{equation}\label{Bohm}
\frac{\delta\Phi({\bf x};T]}{\delta T({\bf x}')}=
\delta^3({\bf x}-{\bf x}')
\left.\frac{\delta S}{\delta\phi({\bf x})}\right|_{\phi=\Phi} .
\end{equation}
It can also be integrated over $d^3x'$ inside an arbitrarily small
region $\sigma_{{\bf x}}$ around ${\bf x}$, to yield
\begin{equation}\label{Bohm_nonloc}
\int_{\sigma_{{\bf x}}} d^3x' 
\frac{\delta\Phi({\bf x};T]}{\delta T({\bf x}')}=   
\left.\frac{\delta S}{\delta\phi({\bf x})}\right|_{\phi=\Phi} .
\end{equation}
Eq.~(\ref{Bohm_nonloc}) is the MFT version of the usual 
single-time Bohmian equation of motion 
$\partial\Phi({\bf x},t)/\partial t=
\delta S/\delta\phi({\bf x})|_{\phi=\Phi}$.
However, (\ref{Bohm}) is more fundamental than (\ref{Bohm_nonloc}) 
because (\ref{Bohm}) does not involve an arbitrary region 
$\sigma_{{\bf x}}$. On the other hand,
the integration inside $\sigma_{{\bf x}}$
is useful for an easier comparison with the conventional
single-time formalism. 
For example,  
from (\ref{Bohm}) and the quantum MFT Hamilton-Jacobi equation
(\ref{HJ}), one finds that $\Phi({\bf x}; T]$
satisfies 
\begin{equation}\label{KGB}
\left[ \left( \int_{\sigma_{{\bf x}}} d^3x'
\frac{\delta}{\delta T({\bf x}')} \right)^2
-\nabla^2_{{\bf x}} +m^2 \right]
\Phi({\bf x};T] 
=-\left. \int_{\sigma_{{\bf x}}} d^3x'
\frac{\delta{\cal Q}({\bf x}';\phi,T]}{\delta\phi({\bf x})}
\right|_{\phi=\Phi}.
\end{equation}
This can be viewed as a MFT Klein-Gordon equation, modified 
with a nonlocal quantum term on the right-hand side. 

The $\delta$-function on the right-hand side of (\ref{Bohm})
implies that the left-hand side of (\ref{Bohm}) 
vanishes when ${\bf x}'\neq {\bf x}$. This implies that 
$\Phi({\bf x};T]$ does not depend on the whole function $T$, 
but only on $T({\bf x})$. Therefore, we have
\begin{equation}\label{locc1}
\Phi({\bf x};T]=\Phi({\bf x},T({\bf x})) .
\end{equation}
Using (\ref{param1}), we can also write
\begin{equation}\label{locc2}
\Phi({\bf x},T({\bf x}))=\Phi({\bf x},x^0)=\Phi(x).   
\end{equation}
Eq.~(\ref{locc1}) shows that $\Phi$ is a MFT object depending 
on many times, with one time $T({\bf x})$ for each 
${\bf x}$. Owing to this MFT property, one does not need a preferred
foliation of spacetime. On the other hand, Eq.~(\ref{locc2}) 
shows that, at the kinematic level, $\Phi$ can also be viewed
as a usual local field in which the MFT nature is not manifest. 

Let us also say a few words on quantum measurements.
The basics of MFT quantum measurements
are described in Sec.~\ref{MFT}, 
while more details on quantum measurements in the Bohmian context 
can be found in 
\cite{bohm2,bohmrep,holbook,nikfpl1}. Here it suffices to say
that the $\chi$-degree of freedom of the measuring apparatus
takes a definite value because $\chi$ is also a 
deterministic degree of freedom obeying a Bohmian equation of motion.
Consequently, owing to (\ref{theta}), the Bohmian equation of motion
for $\phi$ takes a form that it would take if $\Psi[\phi,T]$ were 
equal to one of $\Psi_a$'s in (\ref{expanmeas}). This is how 
the Bohmian interpretation explains the effective collapse 
of the wave functional.   

Now let us give a few notes on the classical limit.
The classical Hamilton-Jacobi equation can also be
formulated as a MFT theory \cite{rov1,rov2}.
The equation of motion can also be written in the MFT form
(\ref{Bohm}). However, in the classical case,
the nonlocal ${\cal Q}$-term in (\ref{HJ}) does not appear,
so that the dynamics is completely local.

\section{Manifestly covariant formulation}
\label{COV}

The MFT formalism was introduced by Tomonaga and Schwinger 
with the intention to provide the manifest covariance of 
QFT in the interaction picture. On the other hand, in this paper 
we are using the Schr\"odinger picture, 
and our presentation in the preceding sections is not manifestly covariant. 
Although the MFT formalism automatically avoids 
the introduction of a preferred foliation of spacetime,  
in the preceding sections time is not treated on an equal footing with 
space. Fortunately, the MFT formalism  
can be formulated in a manifestly covariant way  
\cite{dopl} (see also \cite{rov1,rov2}).
In this section we briefly review the basics of this covariant formalism 
and apply the formalism to the Bohmian interpretation. 

We start by introducing a set of 3 real parameters 
$\{ s^1,s^2,s^3\}\equiv {\bf s}$ that serve as coordinates
on a 3-dimensional manifold. {\it A priori}, 
there is not any relation between these coordinates 
and the spacetime coordinates $x^{\mu}$. 
However, the 3-dimensional manifold can be embedded 
in the 4-dimensional spacetime 
by introducing 4 functions $X^{\mu}({\bf s})$. A 3-dimensional 
hypersurface 
in spacetime is defined by the set of 4 equations
\begin{equation}\label{param2}
x^{\mu}=X^{\mu}({\bf s}) .
\end{equation} 
The 3 parameters $s^i$ can be eliminated, leading to one equation 
of the form $f(x^0,x^1,x^2,x^3)=0$, which, indeed, is an equation 
that determines a 3-dimensional hypersurface in spacetime.
Assuming that the background spacetime metric $g_{\mu\nu}(x)$ is given, the 
induced metric $q_{ij}({\bf s})$ on the hypersurface is
\begin{equation}
q_{ij}({\bf s})=g_{\mu\nu}(X({\bf s}))
\frac{\partial X^{\mu}({\bf s})}{\partial s^i}
\frac{\partial X^{\nu}({\bf s})}{\partial s^j},
\end{equation}
where $X\equiv\{X^{\mu}\}$.
Similarly, a normal to the surface is
\begin{equation}
\tilde{n}_{\mu}({\bf s})=\epsilon_{\mu\alpha\beta\gamma}
\frac{\partial X^{\alpha}}{\partial s^1}
\frac{\partial X^{\beta}}{\partial s^2}
\frac{\partial X^{\gamma}}{\partial s^3}.
\end{equation}
The unit normal transforming as a spacetime vector is
\begin{equation}
n^{\mu}({\bf s})=\frac{g^{\mu\nu}\tilde{n}_{\nu}}
{\sqrt{|g^{\alpha\beta}\tilde{n}_{\alpha}\tilde{n}_{\beta}|}}.
\end{equation} 

Now equations of the preceding sections can be written in a covariant 
form by making the replacements
\begin{equation}\label{covnot}
{\bf x}\rightarrow {\bf s} , \;\;\;\;\;
\frac{\delta}{\delta T({\bf x})} \rightarrow
\frac{\delta}{\delta \tau({\bf s})} \equiv
n^{\mu}({\bf s}) \frac{\delta}{\delta X^{\mu}({\bf s})} .
\end{equation}
The Tomonaga-Schwinger equation (\ref{TS}) becomes
\begin{equation}
\hat{{\cal H}}({\bf s})\Psi[\phi,X]=
in^{\mu}({\bf s}) \frac{\delta\Psi[\phi,X]}{\delta X^{\mu}({\bf s})}.
\end{equation}
For free fields, 
the Hamiltonian-density operator in curved spacetime is
\begin{equation}
\hat{{\cal H}}=\frac{-1}{2|q|^{1/2}}
\frac{\delta^2}{\delta\phi^2({\bf s})}
+\frac{|q|^{1/2}}{2}
[-q^{ij}(\partial_i\phi)(\partial_j\phi)
+m^2\phi^2],
\end{equation}
where $q$ is the determinant of $q_{ij}$.
The Bohmian equation of motion (\ref{Bohm}) becomes
\begin{equation}\label{Bohmc}
\frac{\delta\Phi({\bf s};X]}{\delta\tau({\bf s}')}
=\frac{\delta^3({\bf s}-{\bf s}')}{|q({\bf s})|^{1/2}}
\left.\frac{\delta S}{\delta\phi({\bf s})}\right|_{\phi=\Phi} .
\end{equation}
Similarly, (\ref{KGB}) becomes
\begin{equation}\label{KGBc}
\left[ \left( \int_{\sigma_{{\bf s}}} 
d^3s'\frac{\delta}{\delta\tau({\bf s}')}
\right)^2
+\nabla^i\nabla_i +m^2 \right]
\Phi({\bf s};X] 
=-\int_{\sigma_{{\bf s}}} \frac{d^3s'}{|q({\bf s})|^{1/2}}
\left.\frac{\delta{\cal Q}({\bf s}';\phi,X]}
{\delta\phi({\bf s})} \right|_{\phi=\Phi},
\end{equation}
where
$\nabla_i$ is the covariant derivative with respect to $s^i$
and
\begin{equation}
{\cal Q}({\bf s};\phi,X]=-\frac{1}{|q({\bf s})|^{1/2}}  \frac{1}{2R}
\frac{\delta^2R}{\delta\phi^2({\bf s})}.
\end{equation}

The same hypersurface $\Sigma$ can be parametrized by different 
sets of 4 functions $X^{\mu}({\bf s})$. On the other hand, the quantities 
such as $\Psi[\phi,X]$ and $\Phi({\bf s};X]$ depend on $\Sigma$, 
but do not depend on the way in which $\Sigma$ is parametrized. 
The freedom in choosing 
functions $X^{\mu}({\bf s})$ is a sort of gauge freedom closely related 
to the covariance. To find a solution of the covariant equations 
above, it is convenient to fix a gauge. 
For a timelike hypersurface, the simplest choice of gauge is
\begin{equation}
X^i({\bf s})=s^i .
\end{equation}
This choice implies $\delta X^i({\bf s})=0$, which leads to equations 
similar to those of Secs.~\ref{MFT} and \ref{BOHM}.
For example, (\ref{Bohmc}) becomes
\begin{equation}\label{Bohm2}
(g^{00}({\bf x}))^{1/2} \, \frac{\delta\Phi({\bf x};X^0]}
{\delta X^0({\bf x}')}
=\frac{\delta^3({\bf x}-{\bf x}')}{|q({\bf x})|^{1/2}}
\left.\frac{\delta S}{\delta\phi({\bf x})}\right|_{\phi=\Phi} ,
\end{equation} 
which is the curved-spacetime version of (\ref{Bohm}).

The MFT formalism can also be formulated in a manifestly 
covariant way by introducing a more abstract
formalism \cite{schw} that does not involve parametrizations 
such as (\ref{param1}) or (\ref{param2}).
A functional $F$ of 
the hypersurface is written as $F[\Sigma]$, while the derivative 
operator in (\ref{covnot}) is written as $\delta /\delta\Sigma(x)$.
This derivative is defined as
\begin{equation}
\frac{\delta F[\Sigma]}{\delta\Sigma(x)}=
\lim_{v^{(4)}_x\rightarrow 0} \frac{F[\Sigma']-F[\Sigma]}{v^{(4)}_x},
\end{equation}
where $v^{(4)}_x$ is the 4-volume around $x$ enclosed 
between the hypersurfaces $\Sigma'$ and $\Sigma$. In this language, 
the Bohmian equation of motion (\ref{Bohmc}) can be written in another
covariant form as   
\begin{equation}\label{Bohmc2}  
\frac{\delta\Phi(x;\Sigma]}{\delta \Sigma(x')}
=\frac{\delta^3_{\Sigma}(x-x')}{|q^{}_{\Sigma}(x)|^{1/2}}
\left.\frac{\delta S[\phi,\Sigma]}{\delta_{\Sigma}\phi(x)}
\right|_{\phi=\Phi} ,
\end{equation}
and similarly for other equations.
Here the subscript $\Sigma$ denotes quantities attributed to the
hypersurface $\Sigma$.

\section{Conclusion}
\label{CONCL} 

The MFT formulation of QFT does not involve a preferred foliation of
spacetime, which allows a covariant formulation of QFT.
This naturally leads to a covariant MFT Bohmian interpretation
of quantum fields, which also does not involve a preferred foliation of
spacetime. The covariant dynamics of Bohmian fields does not depend 
on the choice of coordinates. 
When a particular set of 
coordinates is chosen, then the {\em solution} of the MFT Bohmian equation
of motion can be written such that the MFT nature of the field 
is not manifest.
However, the Bohmian {\em equation of motion} itself retains its manifest 
MFT form.
   
\section*{Acknowledgements}
This work was supported by the Ministry of Science and Technology of the
Republic of Croatia under Contract No.~0098002.


\begin{thebibliography}{99}

\bibitem{bohm2}
D.~Bohm, Phys.~Rev.~85 (1952) 180. 
\bibitem{bohmrep}
D.~Bohm, B.J.~Hiley, P.N.~Kaloyerou,
Phys.~Rep.~144 (1987) 321. 
\bibitem{holrep}
P.R.~Holland, Phys.~Rep.~224 (1993) 95. 
\bibitem{holbook}
P.R.~Holland, The Quantum Theory of Motion,
Cambridge University Press, Cambridge, 1993.
\bibitem{nikfpl1}
H.~Nikoli\'c, Found.~Phys.~Lett.~17 (2004) 363. 
\bibitem{gol}
S.~Goldstein, S.~Teufel, quant-ph/9902018.
\bibitem{pin1}
N.~Pinto-Neto, E.S.~Santini, Phys.~Rev.~D 59 (1999) 123517. 
\bibitem{pin2}
N.~Pinto-Neto, E.S.~Santini, Gen.~Rel.~Grav.~34 (2002) 505. 
\bibitem{shoj}
A.~Shojai, F.~Shojai, Class.~Quant.~Grav.~21 (2004) 1. 
\bibitem{bar}
J.A.~deBarros, N.~Pinto-Neto, M.A.~Sagioro-Leal, 
Phys.~Lett.~A 241 (1998) 229. 
\bibitem{col}
R.~Colistete Jr, J.C.~Fabris, N.~Pinto-Neto, 
Phys.~Rev.~D 57 (1998) 4707. 
\bibitem{pn}
N.~Pinto-Neto, R.~Colistete Jr, Phys.~Lett.~A 290 (2001) 219. 
\bibitem{mar}
J.~Marto, P.V.~Moniz, Phys.~Rev.~D 65 (2001) 023516. 
\bibitem{pn2}
N.~Pinto-Neto, E.S.~Santini, Phys.~Lett.~A 315 (2003) 36. 
\bibitem{barb}
G.D.~Barbosa, N.~Pinto-Neto, Phys.~Rev.~D 69 (2004) 065014.
\bibitem{tomo}
S.~Tomonaga, Prog.~Theor.~Phys.~1 (1946) 27.
\bibitem{schw}
J.~Schwinger, Phys.~Rev.~74 (1948) 1439.
\bibitem{dopl}
L.~Doplicher, Phys.~Rev.~D 70 (2004) 064037. 
\bibitem{hort}
G.~Horton, C.~Dewdney, J.~Phys.~A 37 (2004) 11935. 
\bibitem{nikrqm}
H.~Nikoli\'c, quant-ph/0406173,
to appear in Found.~Phys.~Lett.
\bibitem{nikDW}
H.~Nikoli\'c, hep-th/0407228, to appear in
Eur.~Phys.~J.~C.
\bibitem{nikfpl2}
H.~Nikoli\'c,  
Found.~Phys.~Lett.~18 (2005) 123. 
\bibitem{ghose}
P.~Ghose, D.~Home, Phys.~Rev.~A 43 (1991) 6382.
\bibitem{rov1}
C.~Rovelli, gr-qc/0207043.
\bibitem{rov2}
C.~Rovelli, Quantum Gravity,
Cambridge University Press, Camridge, 2004;
http://www.cpt.univ-mrs.fr/$\,\tilde{}\,$rovelli.

\end{thebibliography}
\end{document}